\documentclass[preprint,aps,prd]{revtex4}

\usepackage{epsfig}

\RequirePackage{xspace}
\usepackage{epsfig}
\usepackage{graphicx}
\usepackage{color}
\usepackage{latexsym}
\usepackage{amssymb}

\def\im{\textrm{Im}}

\def\BR{\mathcal{B}}
\def\ACP{\mathcal{A}_{CP}}

\def\mathAzz{A^{00}}

\def\pmathP{\mathcal{P}}
\def\pmathA{\mathcal{A}}
\def\pmathT{\mathcal{T}}

\def\AAij{|A^{ij}|}

\def\AAzz{|A^{00}|}

\def\nn{\nonumber}

\begin{document}

\preprint{}
\title{Extracting the CKM phase angle $\gamma(\equiv \phi_3)$ \\
from isospin analysis in $B\to K \pi$ decays}
\author{C.~S.~Kim}\email{cskim@yonsei.ac.kr} \affiliation{Department
  of Physics, Yonsei University, Seoul 120-749, Korea}
\author{Sechul~Oh}\email{scoh@phya.yonsei.ac.kr} \affiliation{Department
  of Physics, Yonsei University, Seoul 120-749, Korea}
\author{Yeo Woong Yoon}\email{ywyoon@yonsei.ac.kr}\affiliation{Department
  of Physics, Yonsei University, Seoul 120-749, Korea}

\date{\today}
%
%

\begin{abstract}

\noindent We propose a new method to extract the CKM phase angle $\gamma (\equiv \phi_3)$
from the isospin analysis in $B \to K \pi$ decays.
Unlike  previously proposed methods, we do not employ flavor SU(3)
symmetry, so that this method is free from the hadronic uncertainty coming from
the SU(3) breaking effect. Neither we adopt any Dalitz-plot analysis,
which may involve multiple strong phases and large final state interactions.
After including small CP violating terms in $B^+ \to K^0 \pi^+$ and color-suppressed
electroweak penguin contribution in $B^0 \to K^+ \pi^-$, whose values are estimated
from the QCD factorization,
we obtain $\gamma = ({70}^{~+5\, +1\,  +2 }_{-14\, -1\,  -3})^\circ
~~\textrm{or}~~ {106^\circ}< \gamma < {180^\circ}$. The first error is due to
the experimental errors mainly caused by mixing-induced CP asymmetry $S_{{K_s} \pi^0}$.
The second and third errors come from the theoretical uncertainty for two above-mentioned
small contributions, respectively.
Since we utilize only the isospin relations in $B \to K \pi$ decays, this method
will work well, regardless of possible new physics effects unless the isospin
relations do not hold.
\end{abstract}

\maketitle

\section{Introduction}
\label{sec:1}

Investigating validity of the Cabibbo-Kobayashi-Maskawa (CKM) unitary triangle has
been one of the most challenging subjects for testing the standard model (SM).
Great efforts in both experiment at $B$ factories and theoretical side have been
devoted to extract the length of each side and each angle of the triangle.
In this letter, we focus on the CKM angle $\gamma (\equiv \phi_3)$, providing a new
method to extract $\gamma$ by using $B \to K \pi$ decays only.

It was first proposed by Gronau, London and Wyler (GLW) to extract $\gamma$ from
$B \to D K $ decays~\cite{Gronau:1990ra}.
The idea is on the basis of the well-known method using two triangles that are
constructed from the decay amplitude $A(B^+ \to D_+^0 K^+)$ and its charge conjugate
$A(B^- \to D_+^0 K^-)$, respectively, with one common amplitude
$A(B^+ \to \bar{D^0} K^+)=A(B^- \to D^0 K^-)$ in a complex plane.
This method has its own benefit of theoretical cleanness, but it suffers from
some practical difficulty because the triangles are a bit squashed.
Atwood, Dunietz and Soni (ADS)~\cite{Atwood:1996ci} improved the GLW method, and
later the Dalitz analysis was introduced~\cite{Atwood:2000ck}.
These three methods have been used as the most favored methods for extracting
$\gamma$ in experiment.
The current fitting result from UTfit group combining with these three methods
gives $\gamma = 83^\circ \pm 19^\circ$~\cite{UTfit}.

Another promising approach for extracting $\gamma$ is the combined analysis of
$B \to K\pi$ and $B \to \pi\pi$ decays.  As Gronau, Hernandez, London and Rosner
(GHLR) proposed~\cite{Gronau:1994bn}, similarly to the case using $B \to DK$ decays,
one can utilize the decay amplitudes of
$B^\pm \to K^\pm \pi^0$, $B^\pm \to K^0 \pi^\pm$ and $B^\pm \to \pi^\pm \pi^0$ with
the help of flavor SU(3) symmetry.  However, apart from the hadronic uncertainty
coming from the SU(3) breaking effect, the GHLR method had been known
to be spoiled considerably by electroweak (EW) penguin amplitudes~\cite{Deshpande:1994pw}.
Therefore, many authors have elaborated on dealing with the EW penguins and
developed alternative ways to extract $\gamma$ from $B \to K\pi$ decays
\cite{Fleischer:1995cg,Fleischer:1997um,Fleischer:1998bb}.

We would like to propose a new method for extracting $\gamma$ to a good accuracy
by using the well-known {\it isospin relations} in $B \to K\pi$ decays:
\begin{eqnarray}
\label{ISOSPIN-RELATION1}
A(B^0\to K^+\pi^-) - A(B^+\to K^0 \pi^+)
&=&  \sqrt{2} A(B^+\to K^+\pi^0) -\sqrt{2} A(B^0 \to K^0\pi^0),\\
\label{ISOSPIN-RELATION2}
A(\bar B^0\to K^-\pi^+) - A(B^-\to \bar K^0 \pi^-)
&=&  \sqrt{2} A(B^-\to K^-\pi^0) -\sqrt{2} A(\bar B^0 \to \bar K^0\pi^0).
\end{eqnarray}
We do not employ flavor SU(3) symmetry in order to avoid hadronic uncertainties
stemming from the SU(3) breaking effect.
Since we do not impose any theoretical inputs on the $K^+\pi^0$ and $K^0\pi^0$ modes which
are potentially sensitive to new physics effects, this method will work well,
regardless of possible new physics effects unless they spoil the isospin relations.
 We note that significant new physics contribution from the electro-weak penguin diagram may
be present in $B \to K\pi$ decays, as discussed in Ref. \cite{Kim:2007kx}. It has been shown that
current experimental data imply the $r_{_{\rm EW}}$, the ratio of electro-weak penguin amplitude to the strong
penguin amplitude, deviates from the SM expectation value: for example, $r_{_{\rm EW}} = 0.29 \pm 0.13$ whereas the
SM expects $r_{_{\rm EW,SM}}= 0.14 \pm 0.04$.
However, even under the presence of new physics, the $\gamma$ that we measure in the
$B \to K\pi$ decays is the SM $\gamma$ through the {\it re-parametrization invariance}~
\cite{Botella:2005ks} unless new physics comes into the tree diagram.

The isospin relations imply two isospin quadrangles in a complex plane.
In order to extract $\gamma$ from the two isospin quadrangles,
it is crucial to fix the two isospin quadrangles in a common complex plane
using current experimental data.
Belle, BABAR and CLEO collaborations have measured branching ratios (BRs)
and direct CP asymmetries for $B \to K \pi$ decays accurately
as shown in Table I.
Using these data, the length of each side of the two quadrangles could be
determined within $(1 - 5) \%$ error through following definitions,
\begin{eqnarray}
\label{BRs} \BR^{ij} \propto \tau_{B^{(+,0)}}
\frac{{\AAij}^2+\bar{\AAij}^2}{2},~ ~ ~
\ACP^{ij} \equiv
-\frac{{\AAij}^2-\bar{\AAij}^2}{{\AAij}^2+\bar{\AAij}^2},
\end{eqnarray}
where $A^{ij}$ and $\bar A^{ij}$ denote the decay amplitudes of $B \to K^i \pi^j$
and its charge conjugate mode, respectively.

\begin{table}[t]
\caption{Current average data of branching ratios and direct CP asymmetries for
$B \to K \pi$ decays, updated by April 2008 from HFAG \cite{Barberio:2006bi}.}
\label{Table1}
\begin{tabular}{c||c|c}
\hline
decay mode ~~&~~ $\BR(10^{-6})$  ~~&~~ $\ACP$ \\
\hline
$K^0 \pi^+$ ~&~~ $23.1\pm1.0$   ~&~~ $0.009\pm0.025$\\
$K^+ \pi^0$ ~&~~ $12.9\pm0.6$   ~&~ $0.050\pm0.025$\\
$K^+ \pi^-$ ~&~~ $19.4\pm0.6$   ~&~ $-0.097\pm0.012$\\
$K^0 \pi^0$ ~&~~ $9.9\pm0.6$   ~&~~ $-0.14\pm0.11$\\
\hline
\end{tabular}
\end{table}

\section{Extracting $\gamma$ using isospin relation.}
\label{sec:2}

\begin{figure}[t]
\centerline{\epsfig{figure=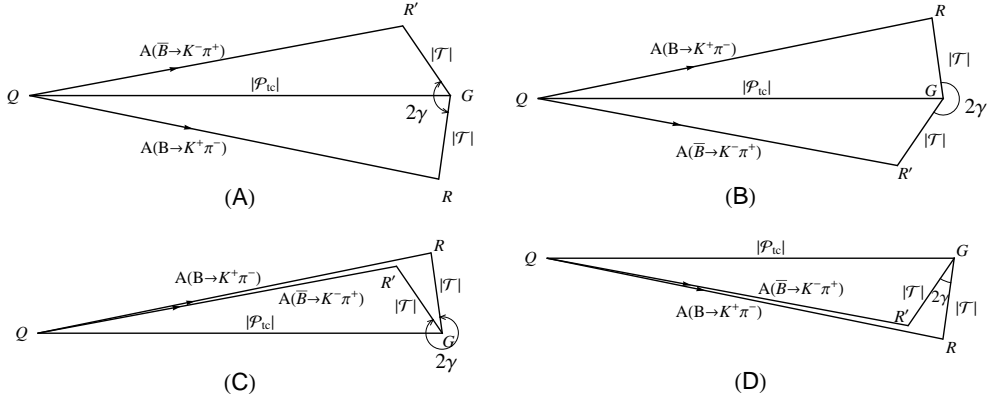, scale=0.32}}
\caption{A four-fold ambiguity arising in constructing two triangles
from Eq. (\ref{AMP-PM}).}
\label{Fig1}
\end{figure}

For a moment, we digress to explain the triangle analysis~\cite{Fleischer:1995cg,BABAR-PHYS}
on determination of the CKM angle $\gamma$.
Applying the quark diagram approach \cite{Gronau:1994bn,Gronau:1995hn} to the $K^0 \pi^+$ decay,
the decay amplitude is parameterized by
$A(B^+\to K^0 \pi^+) = \pmathP_{tc} + (\pmathP_{uc} +
\pmathA) e^{i \gamma}$.
One can roughly estimate that the CP violating terms are very small since
$|\pmathP_{uc}/\pmathP_{tc}| \sim |\pmathA/\pmathP_{tc}|
\sim \mathcal{O}(\lambda^3)$ \cite{Buras:1994pb,Kim:2007kx}.
For the illustration, at first we neglect these small CP violating terms, and
we include them later on. Then approximately,
$A(B^+\to K^0 \pi^+) \simeq A( B^- \to \bar K^0 \pi^-) \simeq \pmathP_{tc}$.
Similarly, the decay amplitudes of $B^0 \to K^+ \pi^-$ and
its CP conjugate mode can be parameterized as
\begin{eqnarray}
\label{AMP-PM}  A(B^0\to K^+ \pi^-) = -\pmathP_{tc} -  \pmathT e^{i \gamma}, ~ ~ ~
 A(\bar B^0\to K^- \pi^+) = -\pmathP_{tc} -
\pmathT e^{- i \gamma},
\end{eqnarray}
where  the small color-suppressed EW penguin contributions are  neglected, as it
has been done in Ref.~\cite{Buras:1995pz}.
However, we will include the contributions  of color-suppressed EW
penguin as well as $(\pmathP_{uc}+\pmathA)/\pmathP_{tc}$ on extracting $\gamma$
in the last part of the analysis. As will be seen, it turns out that these contributions
are minor ones.
Using Eq.~(\ref{AMP-PM}), one can draw two triangles in a complex
plane with the common side $|\pmathP_{tc}|$. Note that this can be done with a four-fold
ambiguity as shown in Fig.~\ref{Fig1} \cite{BABAR-PHYS}. Since we obtain $|\pmathP_{tc}|$,
$|A(B^0 \to K^+ \pi^-)|$ and $|A(\bar {B^0} \to K^- \pi^+)|$ from the current
experimental data, once $|\pmathT|$ is given, one can fix the two triangles with a
four-fold ambiguity.  Then the CKM angle $\gamma$ can be extracted as illustrated in
Fig.~\ref{Fig1}.
Fleischer suggested to make use of flavor SU(3) symmetry in order to estimate $|\pmathT|$
from the branching ratio of $B \to \pi^+ \pi^0$ decay~\cite{Fleischer:1995cg}.
Also, an alternative way to estimate $|\pmathT|$ was proposed by using the factorization
hypothesis~\cite{Fleischer:1997um}. However, it is obvious that both of them would not
be used for a precise measurement of $\gamma$, due to the unpredictable theoretical
uncertainties related to the determination on the value of $|\pmathT|$.
Note that, even with such a theoretical deficiency, this triangle relations look
quite useful and robust due to the very unlikely new physics contributions to $\pmathT$
and $\pmathP_{tc}$.
In our new proposal, we do not bring $|\pmathT|$ into play and, instead, make use of the
isospin relations (\ref{ISOSPIN-RELATION1}) and (\ref{ISOSPIN-RELATION2}) in order to
fix the two triangles in Fig.~\ref{Fig1} \cite{Kim:2007kx}.

It should be noted that
in general four sides and one relative angle between two adjacent sides of a quadrangle
determine the shape of the quadrangle  with  possible discrete ambiguities.
Thus, in order to fix the two
quadrangles in a common complex plane, besides the length of each side of the
quadrangles, one needs at least two additional pieces of information: each on an angle
of the individual isospin quadrangle.  The mixing induced CP asymmetry in $B \to K_S\pi^0$
(denoted by $S_{K_S\pi^0}$) plays an crucial role in our analysis, because it provides
a piece of information on the angle between two sides $A(B^0 \to K^0\pi^0)$ and
$A(\bar B^0 \to \bar K^0\pi^0)$.

\begin{figure}[t]
\centerline{\epsfig{figure=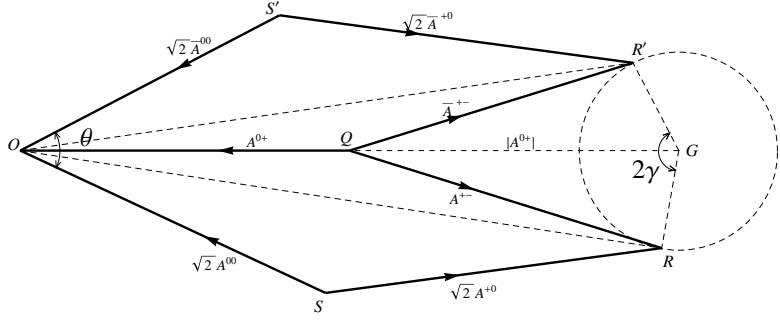, scale=0.50}}
\caption{ Two isospin quadrangles of $B \to K \pi$ decays.  Here two triangles
$\triangle QGR$ and $\triangle QGR^{\prime}$ in the case (A) of Fig.~\ref{Fig1} are
attached to these isospin quadrangles. }
\label{Fig2}
\end{figure}

The two quadrangles can be put together with the common side $|A^{0+}|$ in a complex
plane as shown in Fig.~\ref{Fig2}.
Notice that two triangles $\triangle QGR$ and $\triangle QGR^{\prime}$ in Fig.~\ref{Fig1}
are attached to two isospin quadrangles in Fig.~\ref{Fig2}, which shows the case (A)
of Fig.~\ref{Fig1} as an example.
As the required two additional information,  the angle $\theta$ and the requirement $\overline{GR}$= $\overline{GR'}$
will fix the two isospin quadrangles  in the complex plane with
a certain discrete ambiguity.
Then, the CKM angle $\gamma$ can be extracted from
these two quadrangles.  In other word, one can find the value of $\gamma$ as a function of $\theta$
and extract $\gamma$ from the $\theta$ value given above.

We now relate the angle $\theta$ to the ratio of the decay amplitudes as
\begin{equation}
\label{DEF-THETA}
\theta  \equiv  (\textrm{sign})~ \angle SOS^\prime = \arg \left( \frac{\bar\mathAzz}{\mathAzz} \right),
\end{equation}
where the sign is ``plus'' for the case of $\overline{O S^\prime}$ upper than
$\overline{OS}$ and ``minus'' for the opposite case.  The crucial point is that
$\theta$ can be obtained from the mixing induced CP asymmetry $S_{K_S\pi^0}$ of
$B^0 \to K_S\pi^0$ decay.  It can be easily seen from the expression of $S_{K_S\pi^0}$:
\begin{equation}
\label{EXPLICIT-SZZ} S_{K_S\pi^0}= -\frac{2 \AAzz \bar{
\AAzz}}{{\AAzz}^2 + \bar {\AAzz}^2} \im ( e^{i \theta} e^{- 2 i
\beta}).
\end{equation}
We use  $2 \beta = 42.7^\circ \pm 2.0^\circ$~\cite{Barberio:2006bi} that is averaged
over the mixing induced CP asymmetries of $b \to c \bar{c} s$ processes.  Thanks to
the recent BABAR measurement, the error of $S_{K_S\pi^0}$ is a bit reduced, resulting
in $S_{K_S\pi^0}=0.38\pm0.19$~\cite{Barberio:2006bi}.
Using this experimental result, we determine the value of $\theta$ from
Eq.~(\ref{EXPLICIT-SZZ}) as
 $\theta = 20^\circ \pm 12^\circ~~\textrm{or}~
-115^\circ \pm 12^\circ.$

\begin{figure}[t]
\centerline{\epsfig{figure=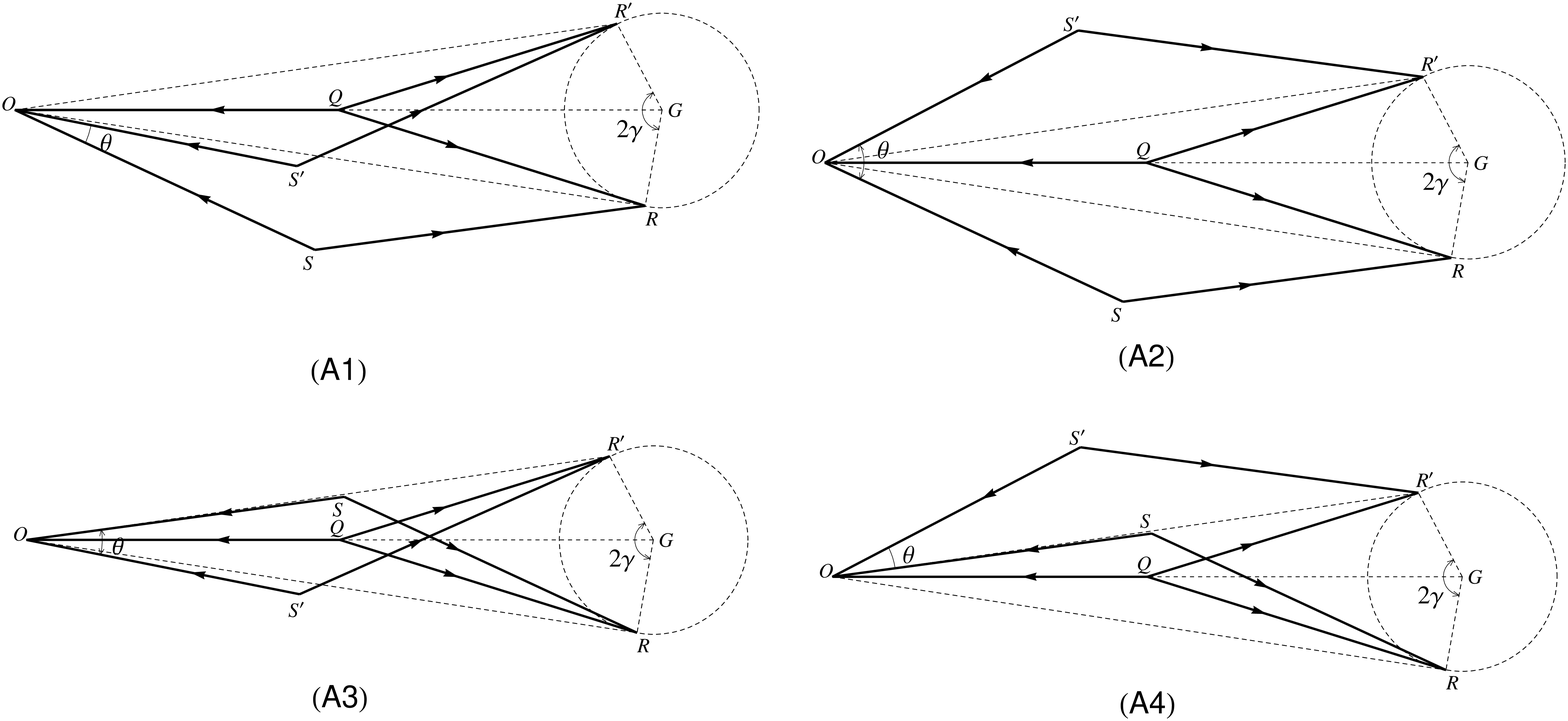, scale=0.25}}
\caption{ A four-fold ambiguity arising in constructing two isospin quadrangles of $B \to K
\pi$ decays, which corresponds to the case (A) of Fig.~\ref{Fig1}.  }
\label{Fig3}
\end{figure}

Next, let us discuss the discrete ambiguity involved in extracting $\gamma$.  We recall
that a four-fold ambiguity is involved in constructing $\triangle QGR$ and
$\triangle QGR^{\prime}$ as shown in Fig.~\ref{Fig1}.
For two triangles fixed as in the case (A) of Fig.~\ref{Fig1}, a two-fold ambiguity
arises when one constructs each isospin quadrangle (attached to each triangle),
depending on whether the position of the remaining apex $S(S^{\prime})$ is upper than
or lower than $\overline{OR}(\overline{OR^{\prime}})$.
Thus, another four-fold ambiguity arises when the two isospin quadrangles are constructed
for the two fixed triangles. In Fig.~\ref{Fig3} this four-fold ambiguity is depicted
as (A1), $\cdots$, (A4) for the case (A).
Therefore, there appears a sixteen-fold ambiguity in total, which is called
(A1), (A2), $\cdots$, (D4).
Accordingly, there are sixteen distinct $\gamma$'s as a function of $\theta$.

\begin{figure}[b]
\centerline{\epsfig{figure=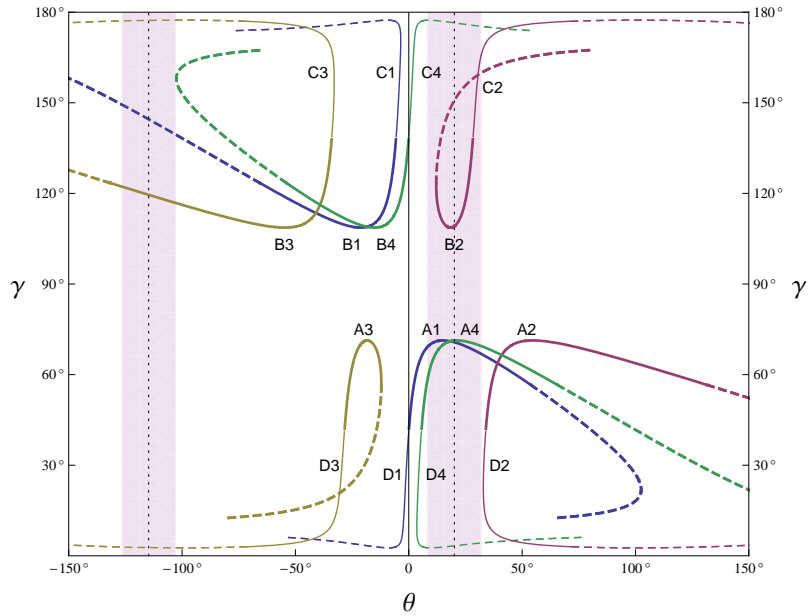, scale=0.6}}
\caption{Plot of $\gamma$ versus $\theta$ for sixteen distinct cases. Each curve is
obtained from the central values of branching ratios and direct CP asymmetries.
The extracted values of $\theta$  is represented
by the shaded region.  The dashed lines imply the region where $|\pmathT /
\pmathP_{tc} | \geq 1$.}
\label{Fig4}
\end{figure}

Each plot of $\gamma$ versus $\theta$ is shown in Fig.~\ref{Fig4}.
As this figure displays, one can extract the possible values of $\gamma$ using the value of
$\theta$. The result within $1 ~\sigma$ is
\begin{eqnarray}
\label{GAMMA-EXT1}
A1&:&~~ \gamma = ({71}^{+0~+4}_{-4~-4})^\circ,\nn \\
\label{GAMMA-EXT2}
A4&:&~~ \gamma = ({71}^{~+0~+5}_{-13~-6})^\circ,\nn \\
\label{GAMMA-EXT3}
B2 ~\textrm{and} ~ C2 &:&~~ \gamma = ({109}^{+57+66}_{-0~-3})^\circ, \nn \\
\label{GAMMA-EXT4}
B3&:&~~ \gamma = ({119}^{+2~+2}_{-3~-2})^\circ\,,
\end{eqnarray}
where the first errors are due to the error of $S_{K_S \pi^0}$, while the second errors
come from the combined error of all branching ratios and direct CP asymmetries
in quadrature.
Here we have discarded the region where $|\pmathT / \pmathP_{tc} | \geq 1$,
which is much more conservative compared to the SM estimation of
$0.18\pm0.11$ for QCDF \cite{Beneke:2001ev} and $0.15\pm0.12$
for PQCD \cite{Li:2005kt} within 2$\sigma$ range.
We note that the $\gamma$ solutions for the $B2$ and $C2$ case are very sensitive
to every experimental error.
Consequently, the estimation of $\gamma$ for the region greater than $90^\circ$
is poor in our method.
Combining above all the results, 
\begin{eqnarray}
\label{eq:gam-neg}
\gamma = {71^\circ}^{+5^\circ}_{-14^\circ}~ [0^\circ < \gamma < 83^\circ ]
~~\textrm{or}~~ 106^\circ < \gamma < 180^\circ ~[106^\circ < \gamma < 180^\circ ]\,,
\end{eqnarray}
where the values in brackets show $2\sigma$ result.
It should be emphasized that, as can be seen from Fig.~\ref{Fig4}, more accurate
measurement of $S_{K_S\pi^0}$ as well as branching rations and direct CP
asymmetries will lead to more precise estimation of $\gamma$ especially for the region
less than $90^\circ$. One can find the maximum value of $\gamma$ for the cases (A) and (D):
$\gamma \leq 71^\circ$, and the minimum value of $\gamma$ for the cases
(B) and (C): $\gamma \geq 109^\circ$, from Fig.~\ref{Fig4}. These bounds on $\gamma$ are
sharply consistent with the Fleischer-Mannel bound \cite{Fleischer:1997um},
$ 0^\circ < \gamma < \gamma_0 ~~\bigvee~~ 180^\circ - \gamma_0 < \gamma < 180^\circ$, where
$\gamma_0$ can be obtained to be $72^\circ$.

As we mentioned, the result of Eq.(\ref{eq:gam-neg}) is obtained with neglecting
the small contributions, $(\pmathP_{uc}+\pmathA)/\pmathP_{tc} \equiv \varepsilon_a e^{i\phi_a}$
and the ratio of the color-suppressed EW penguin to the strong penguin
$\pmathP^C_{EW}/\pmathP_{tc} \equiv  \varepsilon_C e^{i\phi_C}$.
Now we include these contribution, using the theoretical estimation in the
framework of QCD Factorization. The authors of Ref.~\cite{Beneke:2001ev} calculated that
$\varepsilon_a = 0.02 \pm 0.004,~ \phi_a = 13.6^\circ \pm 4.4^\circ $,
$\varepsilon_C = 0.017 \pm 0.011,~ \phi_a = -67.7^\circ \pm 49.7^\circ $.
We scan these four parameters within $1\sigma$ variation of the estimation. And
we followed the method that we explained above in order to get the solution of $\gamma$ under
the consideration of each scanned parameter values.
Then we find the variation of $\gamma$
values according to the variation of the four parameters.
Since $\ACP(K^0\pi^+) \approx 2 \varepsilon_a
\sin\gamma \sin\phi_a$, the current experimental data for $\ACP(K^0\pi^+)$ strongly
constrain the parameter region of $\varepsilon_a$ and $\phi_a$. We discard the
parameter values that go beyond this constraint.
Then, following result is obtained: 
\begin{equation}
\label{gam-sol}
\gamma = ({70}^{~+5\, +1\,  +2 }_{-14\, -1\,  -3})^\circ~[0^\circ < \gamma < 80^\circ ]
~~\textrm{or}~~ {106^\circ}< \gamma < {180^\circ} ~[104^\circ < \gamma < 180^\circ ]\, ,
\end{equation}
where the values in brackets show $2\sigma$ result.
The first error is due to the experimental errors mainly caused by mixing-induced
CP asymmetry $S_{{K_s} \pi^0}$. The second and third errors come from the theoretical
uncertainty of $\varepsilon_a e^{i\phi_a}$ and $\varepsilon_C e^{i\phi_C}$, respectively.
Comparing the central value and the errors with Eq.(\ref{eq:gam-neg}), we can see that
the small parameters
$\varepsilon_a e^{i\delta_a}$ and $\varepsilon_Ce^{i\delta_C}$ are not significant for
extracting $\gamma$ in this method.

\section{Conclusion.}
\label{sec:3}

We have presented a new method for extracting the CKM phase $\gamma$
using the isospin analysis in $B \to K\pi$ decays.  In this method, flavor SU(3)
symmetry is not employed so that the hadronic uncertainty arising from the SU(3) breaking effect
does not spoil the method. Neither we adopt any Dalitz-plot analysis, which may involve multiple
strong phases and large final state interactions.
Since we utilize only the isospin relations in $B \to K \pi$ decays, this method
will work well, regardless of possible new physics effects unless the isospin
relations do not hold.
Using the current data for $B \to K\pi$ including
$S_{K_S \pi^0}$, we have found that
$\gamma = ({70}^{~+5\, +1\,  +2 }_{-14\, -1\,  -3})^\circ
~~\textrm{or}~~ {106^\circ}< \gamma < {180^\circ}$ at $1\sigma$ level,
where the first error is due to the experimental errors,
and the second and the third errors come from the theoretical
uncertainty of the CP violating terms
in $B^+ \to K^0 \pi^+$ and color-suppressed EW penguin term, respectively.
\\



\vspace{1cm} \centerline{\bf ACKNOWLEDGMENTS}

\noindent The work of C.S.K. was supported in part by CHEP-SRC and
in part by the KRF Grant funded by the Korean Government (MOEHRD)
No. KRF-2005-070-C00030. The work of S.O. was supported by the
Second Stage of Brain Korea 21 Project. The work of Y.W.Y. was
supported by the KRF Grant funded by the Korean Government (MOEHRD)
No. KRF-2005-070-C00030.



\end{document}